\pdfoutput=1
\documentclass[11pt]{article}
\usepackage[margin=1in]{geometry}
\usepackage{newtxtext,newtxmath}

\usepackage{booktabs}
\usepackage{amsmath}
\usepackage{graphicx}
\usepackage{siunitx}
\usepackage{url}
\usepackage{placeins}
\usepackage{tikz}
\usetikzlibrary{arrows.meta, positioning, shapes.geometric, fit, calc}
\newcommand{\NA}{\multicolumn{1}{c}{--}}
\makeatletter
\newcommand\unnumberedsection[1]{\section*{#1}}
\renewcommand\thebibliography[1]{%
  \unnumberedsection{References}\@mkboth{}{}%
  \list{[\arabic{enumi}]}%
    {\settowidth\labelwidth{[#1]}\leftmargin\labelwidth
     \advance\leftmargin\labelsep
     \usecounter{enumi}}%
  \def\newblock{\hskip .11em plus .33em minus .07em}%
  \sloppy\clubpenalty4000\widowpenalty4000
  \sfcode`\.=1000\relax}
\makeatother

\title{Deployment-constrained key estimation with a DSP-guided transformer reranker}

\author{Pardis Noorzad\\[-0.2ex]
{\small Cue Labs}\\[-0.2ex]
{\small pardis@cuelabs.ca}}
\date{}

\begin{document}

\maketitle
\begingroup
\renewcommand\thefootnote{}
\footnotetext{This work has been submitted to the IEEE for possible publication.
Copyright may be transferred without notice, after which this version may no
longer be accessible.}
\endgroup

\begin{abstract}
Local music applications estimate key during library import and benefit from an
analyzer that is small, fast, and runs in-process. We start from a compact
chroma-template key analyzer that fits this budget but lags behind learned
references.
Learned systems reach higher accuracy, but their installed model and runtime
dependencies can occupy tens to hundreds of megabytes. We present key lime, a
21.5k-parameter DSP-guided transformer reranker with a 115~KB model artifact
and a 76.7~MB installed footprint, including the extracted ONNX Runtime. It
adds 0.32~ms per track to deterministic DSP scoring. Trained from labels on
4{,}040 FMAK tracks, with 449 tracks for validation and 1{,}000 disjoint FMAKv2
tracks held out for evaluation, it improves MIREX from 61.08 to
$73.55 \pm 0.46$ and exact accuracy from 52.30 to $67.07 \pm 0.65$ over three
seeds (mean $\pm$ population SD). It approaches a same-manifest local madmom
KeyCNN reference at 74.16 MIREX. A 12-parameter circular input stage matches
the accuracy range of a 468-parameter linear input projection.
\end{abstract}


\section{Introduction}

Local music applications such as
Cueport\footnote{\url{https://usecueport.com}} display estimated musical keys
during library import. Because this analysis runs in-process on the user's
laptop~\cite{kleppmann2019localfirst}, the estimator's footprint matters as
much as its accuracy: it must add little binary size and few runtime
dependencies, and finish fast enough that importing thousands of tracks remains
practical. Compact chroma-template
methods~\cite{krumhansl1982tonal,temperley1999key,faraldo2016key} meet this
budget. Our Rust analyzer, trail mix\footnote{\url{https://github.com/djpardis/trailmix/releases/tag/key-lime-paper-v1}},
computes tempo, beat-position, key, and waveform summaries locally; only key is
evaluated here. Its key scorer reaches 61.08 MIREX on our 1{,}000-track FMAKv2
holdout, compared with 74.16 for local madmom
KeyCNN~\cite{korzeniowski2018keycnn}. S-KEY~\cite{kong2025skey} reports 72.1
and 73.2 MIREX on the full 5{,}489-track protocol. This key gap motivates
adding a small learned component while retaining the broader DSP analyzer.
We call the deterministic trail mix key scorer KL-DSP and the learned reranker
KL-FMAK.
Madmom and S-KEY occupy 174~MB and 436~MB when installed
(Table~\ref{tab:deploy}).

We introduce key lime, a transformer with 21{,}528 parameters, exported as a
115~KB ONNX file that adds 0.32~ms per track to DSP scoring. It reranks the 24
DSP key scores using frame-level chroma context. The selected model uses an
identity-initialized DSP-score projection and a circular input convolution as
low-cost structural choices. The ablation in Table~\ref{tab:ablation} shows
that the frame-sequence branch accounts for the measured accuracy gain. Circular
input and identity initialization were not distinguishable at three seeds.
The circular input shares weights over pitch classes, giving the input stage a
transposition-equivariant form related in spirit to the transposition
objectives of STONE and MT\textsuperscript{2}~\cite{kong2024stone,kong2025mt2}. Training uses
4{,}040 labeled FMAK tracks, selects checkpoints on 449 disjoint FMAK
validation tracks, and reports final and diagnostic results on a sealed
1{,}000-track FMAKv2 holdout.

Over three seeds, KL-FMAK raises held-out MIREX from 61.08 to
$73.55 \pm 0.46$, approaching the local madmom reference at 74.16 on the same
files. Its
76.7~MB installed footprint, dominated by ONNX Runtime, is 2.3$\times$
smaller than madmom's and 5.7$\times$ smaller than S-KEY's; the model file
itself is 115~KB. Published S-KEY and madmom scores use the full 5{,}489-track
protocol and provide context for the local same-manifest results.

Our contributions are: (i) a compact DSP-guided transformer reranker for a DSP
key scorer; (ii) a held-out,
three-seed evaluation against DSP and local madmom with paired-bootstrap and
McNemar tests; and (iii) an ablation separating the sequence branch, circular
input, identity initialization, and distillation, together with measured model,
runtime, and installed footprints.

\section{Related Work}

\subsection{Key estimation: DSP and learned approaches}

Major and minor keys produce characteristic pitch-class distributions that
remain consistent across instruments; comparing a chroma histogram against
24 fixed key templates via Pearson correlation recovers the key using
deterministic scoring in the Krumhansl-Schmuckler
tradition~\cite{krumhansl1990cognitive}. Profiles from
Krumhansl-Kessler~\cite{krumhansl1982tonal} and
Temperley~\cite{temperley1999key} target common-practice tonality;
genre-adapted variants such as Faraldo's EDM
templates~\cite{faraldo2016key} address the distinct harmonic vocabulary
of electronic dance music.

CNN-based methods replace hand-crafted profiles with filters learned from
labeled data. madmom KeyCNN~\cite{korzeniowski2018keycnn} trains a
convolutional network on labeled chroma frames; a CNN captures richer
correlations among pitch classes than a fixed template, raising accuracy
above profile-matching across diverse genres at the cost of a learned
model artifact and inference runtime.

Self-supervised methods reduce dependence on labeled corpora by pre-training on
large unlabeled collections. STONE~\cite{kong2024stone} and its successor
S-KEY~\cite{kong2025skey} exploit pitch transposition as a free
supervisory signal: the same audio shifted by $n$ semitones should yield
a representation shifted by $n$ in key space, and training this
consistency at scale via cross-power spectral density losses reduces reliance
on labeled data. STONE variants use limited supervision, whereas S-KEY requires
no human annotation; MT\textsuperscript{2} combines CPSD
with a contrastive objective~\cite{kong2025mt2}. These systems need an external
inference runtime, which can dominate their installed footprint.

\subsection{Existing key-analysis tools}

Essentia~\cite{bogdanov2013essentia},
libKeyFinder\footnote{\url{https://github.com/mixxxdj/libkeyfinder}}, and
stratum-dsp\footnote{\url{https://crates.io/crates/stratum-dsp}} provide local
key analysis in C++ or Rust. We include them in the footprint comparison
because they represent the deployment profile of local DSP analyzers without a
separate learned model runtime. The method studied here is a compact learned
global-key DSP-score reranker that preserves much of that local deployment
profile while correcting some key-estimation errors. It adds a 115~KB model
and an ONNX Runtime dependency to a local analyzer rather than replacing the
analyzer with a larger learned system.
We use trail mix as the base scorer because it is the target local analyzer and
already computes tempo, beat-position, key, waveform summaries, and the
24-key score vector consumed by the reranker. We report Essentia's EDMM profile
because it is the EDM-oriented profile rather than selecting an Essentia profile
on the holdout. The libKeyFinder result shows that another compact DSP scorer
is stronger than KL-DSP on this holdout; reranking other scorers' 24-key score
vectors is future work.

\section{Method}

\subsection{DSP key scorer}
\label{sec:key}

The key analyzer computes chroma from two complementary front ends, a
fixed-window Goertzel bank and a Constant-Q
transform~\cite{brown1991cqt,schorkhuber2010cqt}, and normalizes and
averages them into $\tilde c$. Scoring uses Pearson correlation
$\rho(\tilde c, p_k)$ against 24 key templates $p_k$ (12 major, 12 minor)
drawn from Krumhansl-Kessler~\cite{krumhansl1982tonal},
Temperley~\cite{temperley1999key}, and Faraldo EDM
profiles~\cite{faraldo2016key}, plus a GiantSteps Key-fitted profile in the
archived scorer. That fitted profile was learned from trail mix GiantSteps Key
benchmark output and does not use FMAK or FMAKv2 audio or labels. Correlations
are averaged across the major/minor template pairs from each profile family.
The winning template sets \texttt{global\_key}, and the reported
\texttt{confidence} is the margin between the top and second-best correlations.
This paper evaluates only the global key estimate.

\subsection{Compact learned key backend}
\label{sec:learned-key}

The learned key backend is a compact transformer
encoder~\cite{vaswani2017attention} over the same frame-level chroma
sequence used by the DSP key scorer. The KL-FMAK artifact uses one encoder
layer, $d_m{=}36$, four attention heads, a 72-dim feed-forward block,
dropout~0.3~\cite{srivastava2014dropout}, and weight
decay~0.05~\cite{loshchilov2019adamw}. Training uses AdamW with learning rate
0.001, batch size 16, at most 25 epochs, validation-based early stopping, class
balancing over the 24 labels, a five-epoch DSP-margin curriculum, and
MIREX-aware soft targets that spread mass $\lambda{=}0.2$ to fifth, relative,
and parallel neighbours of the true key. Transposition augmentation rotates each
input sequence, label, and DSP prior consistently through all 12 semitones.
Inputs use the first 256 chroma frames, which covers about 32 seconds at trail
mix's chroma rate of approximately 8~Hz; shorter tracks are zero-padded to the
same length. The exported model uses no padding mask, so mean pooling includes
those zero-padded frames exactly as in training. Early stopping uses validation
MIREX. Section~\ref{sec:data} defines the train, validation, and holdout split
used for these settings.
The DSP-margin curriculum orders early training examples by the DSP score
margin between the top two candidate keys, starting with high-margin examples
and linearly exposing the full training set over five epochs.

\begin{figure}[!ht]
\centering
\includegraphics[width=.48\textwidth]{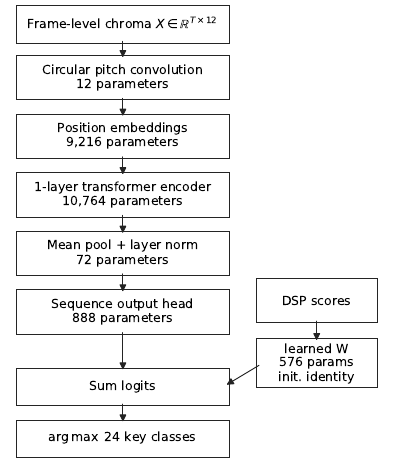}
\caption{KL-FMAK architecture and trainable parameter split. The circular
convolution is part of the selected KL-FMAK model. The six blocks sum to
21{,}528 parameters in a 115~KB fp32 ONNX file.}
\label{fig:architecture}
\end{figure}

Figure~\ref{fig:architecture} summarizes the two input branches and trainable
parameter split.
The DSP-guided reranker fuses the frame sequence with the 24-dim DSP key
score vector $s_{\text{dsp}}$ emitted over the 24 templates. This vector is the
raw averaged Pearson-correlation score vector from the full-track aggregate
chroma; it is not normalized before the identity-initialized projection.
Writing $z_{\text{seq}}\in\mathbb{R}^{24}$ for the pooled encoder logits, the
reranker logits and prediction are
\begin{equation}
z = z_{\text{seq}} + W\,s_{\text{dsp}},\qquad
\hat{y} = \arg\max_{k} z_k,
\label{eq:residual}
\end{equation}
with trainable $W\in\mathbb{R}^{24\times24}$ initialized to the identity, so
before training the DSP score vector is already present in the logits. The
encoder then learns a residual correction over the DSP prediction. This
matches the deployment setting: the local analyzer already computes a useful
24-key DSP ranking, and the learned backend is intended to correct that ranking
under the small-footprint budget rather than replace the analyzer with a
standalone classifier.
This residual form also gives the small model an easier target than a
from-scratch global-key classifier: it can preserve high-confidence DSP
decisions and spend its limited capacity on frames where chroma context changes
the ranking. The learn-only-$W$ ablation in Table~\ref{tab:ablation} isolates
the DSP-score projection and falls below the DSP baseline, so the gain is not a
learned lookup over DSP scores alone. The frame sequence is the evidence source
that lets the residual branch improve the template ranking.

KL-FMAK replaces the linear input projection with a circular 1-D convolution
$g$ over the 12 pitch classes that shares one kernel across all positions. The
single chroma channel feeds three shared output channels with a three-bin
kernel, producing $12\times3=36$ features per frame. For a semitone rotation
$R_t$, a cyclic shift of the pitch-class axis applied to the input,
\begin{equation}
g(R_t\,x) = R_t\,g(x),
\label{eq:equivariant}
\end{equation}
so transposition equivariance is built into this input stage. STONE and
MT\textsuperscript{2} use CPSD losses for transposition
objectives~\cite{kong2024stone,kong2025mt2}. The rest of the model is
unconstrained, so
the guarantee applies to the input stage. The motivation was to share
input-stage statistics across pitch transpositions, so limited training data
need not learn separate first-layer patterns for each tonic. Training minimizes
MIREX-aware soft-target cross-entropy,
\begin{equation}
\mathcal{L} = \mathrm{CE}(q,p),
\label{eq:loss}
\end{equation}
where $p=\mathrm{softmax}(z)$ and the target
$q=(1-\lambda)e_{y}+\lambda r$ places residual mass $\lambda$ on the fifth,
relative, and parallel neighbours of the true key $y$ in proportion to their
MIREX weight. The fifth relation includes both same-mode fifth directions,
while the relative relation follows the directional MIREX convention. We use
$\lambda{=}0.2$ and a three-bin pitch convolution kernel.
KL-FMAK trains on 4{,}040 FMAK tracks and selects checkpoints on a deterministic
449-track FMAK validation split. An ablation distills a frozen madmom KeyCNN
with $\alpha{=}0.3$ and temperature $T{=}4$~\cite{hinton2015distilling}, but
the final model uses labels only.

\section{Evaluation}

\subsection{Data split and metrics}
\label{sec:data}

FMAKv2 provides key annotations for 5{,}489 Free Music Archive
tracks~\cite{wong2024fmakv2,wong2023fmak,defferrard2017fma}. We seal a
deterministic 1{,}000-track holdout (fingerprint
\texttt{43a35a0b...c863}) before model selection. After removing those
track IDs, 4{,}489 FMAK tracks remain; a deterministic hash split assigns 4{,}040 to
training and 449 to validation. Track ID overlap is zero. The training pool
uses FMAK labels while the holdout uses FMAKv2 labels; FMAKv2 revises about
200 annotations, so the label sources are not identical. The 61.08 MIREX DSP
baseline is regenerated from the archived source revision on this sealed
manifest.
The holdout ablations and sensitivity checks below are reported as diagnostics
rather than as validation criteria.

Key uses the MIREX weighted score: 1.0 exact, 0.5 perfect fifth, 0.3 relative,
0.2 parallel, and 0 otherwise. Paired comparisons report a bootstrap 95\%
confidence interval on the per-track MIREX delta~\cite{efron1986bootstrap}.
For KL-FMAK versus deterministic baselines, each track's MIREX score is
averaged across the three KL-FMAK seeds, the deterministic baseline score for
that same track is subtracted, and the bootstrap resamples those 1{,}000
per-track deltas.
McNemar tests are exact-binomial tests on paired exact-match outcomes, not on
the weighted MIREX scores~\cite{mcnemar1947test}.

\subsection{Results and ablations}

Table~\ref{tab:key} separates published reference values from local
same-manifest evaluations. The published S-KEY and madmom values use all
5{,}489 tracks~\cite{kong2025skey}; local rows use only the sealed holdout. The
KL-FMAK row in Table~\ref{tab:key} is the labels-only final variant reported in
Table~\ref{tab:ablation}. KL-FMAK improves over KL-DSP by $12.47$ MIREX and
$14.77$ exact points on average. The representative seed 1729 reaches 73.53
MIREX with a paired
95\% CI of $[+9.51,+15.40]$ points and exact-match McNemar
$p=9.4\times10^{-16}$. The three-seed mean lies 0.61 MIREX below local
madmom, while its exact accuracy is 0.37 points higher.
The three KL-FMAK seed MIREX values are 73.00, 73.53, and 74.12; seed 1729 is
the selected representative artifact because it is closest to the three-seed mean.
The local madmom comparison is a deterministic same-manifest point estimate.
A seed-averaged paired bootstrap gives KL-FMAK minus madmom as $-0.61$ MIREX
with 95\% CI $[-3.27,+2.03]$, so the point gap is not resolved by this sample.
A single-seed sensitivity check around the soft-target mass gives MIREX/exact
of 72.65/65.70 for $\lambda{=}0.1$, 73.53/67.00 for $\lambda{=}0.2$, and
73.26/66.80 for $\lambda{=}0.3$ (seed 1729; cf. the 73.55 three-seed mean for
the same $\lambda{=}0.2$ recipe), so this setting does not reveal a sharp
optimum.

Against KL-DSP, the selected seed converts 244 of KL-DSP's 477 non-exact
tracks into exact matches: 59 fifth errors, 42 relative errors, 34 parallel
errors, and 109 other errors. The same KL-FMAK seed's final non-exact
distribution is 76 fifth, 51 relative, 60 parallel, and 143 other. This
distribution is over KL-FMAK predictions, not the remaining subset of KL-DSP's
original non-exact tracks. Net, the reranker corrects more DSP errors than it
introduces: it converts 244 non-exact tracks to exact against 97 demotions of
previously exact tracks, the $+147$-track net that raises exact accuracy from
52.30\% to 67.00\% for this seed. The demotions are 19 fifth, 18 relative,
22 parallel, and 38 other errors, so they are not confined to one harmonic
relation.
For deployment settings that prefer avoiding demotions over maximizing the
average exact rate, the relevant next evaluation is a confidence-gated deferral
rule that keeps DSP predictions on selected tracks; the present study reports
the ungated reranker.

\begin{table}[t]
\centering
\caption{FMAKv2 key estimation, MIREX / exact (\%). Published S-KEY and
madmom rows are quoted from Kong et al.~\cite{kong2025skey} and use the full
5{,}489-track protocol. Local rows use the same 1{,}000-track holdout;
KL-FMAK reports mean $\pm$ population standard deviation over three seeds. A
dash marks a quantity not reported by the cited source, or an SD that is not
applicable for a single deterministic run.}
\label{tab:key}
\vspace{4pt}
\small
\setlength{\tabcolsep}{1.8pt}
\begin{tabular*}{\columnwidth}{@{\extracolsep{\fill}}l
  S[table-format=2.2]
  S[table-format=1.2]
  S[table-format=2.2]
  S[table-format=1.2]@{}}
\toprule
& \multicolumn{2}{c}{MIREX (\%)} & \multicolumn{2}{c}{Exact (\%)}\\
\cmidrule(lr){2-3}\cmidrule(l){4-5}
System & {Mean} & {SD} & {Mean} & {SD}\\
\midrule
S-KEY 60k (pub.) & 72.10 & \NA & \NA & \NA\\
S-KEY 1M (pub.) & 73.20 & \NA & \NA & \NA\\
madmom (pub.) & 73.10 & \NA & \NA & \NA\\
\midrule
KL-DSP & 61.08 & \NA & 52.30 & \NA\\
KL-FMAK (labels-only, final) & 73.55 & 0.46 & 67.07 & 0.65\\
madmom (local) & 74.16 & \NA & 66.70 & \NA\\
\bottomrule
\end{tabular*}
\end{table}

Table~\ref{tab:ablation} shows that access to the frame sequence is necessary:
the learn-only-$W$ row trains only the DSP-score projection with
$z_{\text{seq}}{=}0$ and remains below DSP. This ablation tests access to
sequence evidence and DSP scores, not attention against every possible
frame-sequence architecture. The three-seed mean is unchanged by circular input
and identity initialization;
random initialization is numerically similar to identity initialization.
The circular input stage uses 12 trainable weights, compared with 468 for the
linear input projection (468 linear versus 12 circular; net +456 parameters in
Table~\ref{tab:ablation}). The full model therefore keeps the linear-input
variant's accuracy range while using about 39$\times$ fewer input-stage
parameters.
Labels-only training is numerically above the distilled row. The 1.05-point gap
is small relative to the three-seed variation. A seed-matched paired bootstrap
over track IDs
gives labels-only minus distilled full as $+1.05$ MIREX with 95\% CI
$[-0.43,+2.51]$. Labels-only exceeds learn-only-$W$ by $+13.30$ MIREX with
95\% CI $[+10.67,+16.01]$. Removing the DSP-margin curriculum gives
73.16$\pm$0.62 MIREX, close to the final labels-only row. The reported gain is
present in labels-only training, so distillation is optional. The learn-only-$W$
row starts from the DSP score vector, but training updates $W$ under FMAK labels,
class weighting, and soft targets; validation early stopping therefore need not
preserve the identity map on the FMAKv2 holdout. Inspecting the
selected seed-1729 $W$ shows learned structure rather than an identity map.
Diagonal weights average 0.80, mean absolute off-diagonal weight is 0.225, and
fifth/relative/parallel entries average $+0.19/+0.18/+0.36$, while other
off-diagonal entries average $-0.12$. This raises musically adjacent
alternatives but is insufficient without frame evidence, matching the
learn-only-$W$ ablation.

\begin{table}[t]
\centering
\caption{FMAKv2 ablation, MIREX / exact (\%), mean $\pm$ population standard
deviation over three seed-matched runs on the sealed holdout. Recipe and
architecture rows use labels-only training unless stated otherwise.
$\Delta$DSP is MIREX points relative to the 61.08 KL-DSP baseline. The table
first compares training objectives, then recipe and architecture variants.}
\label{tab:ablation}
\vspace{4pt}
\small
\setlength{\tabcolsep}{1.6pt}
\begin{tabular*}{\columnwidth}{@{\extracolsep{\fill}}l
  S[table-format=5.0]
  S[table-format=2.2]@{\,$\pm$\,}
  S[table-format=1.2]
  c
  S[table-format=2.2]
  S[table-format=2.2]@{\,$\pm$\,}
  S[table-format=1.2]@{}}
\toprule
Variant & {Params} & \multicolumn{2}{c}{MIREX (\%)} &
  \multicolumn{2}{c}{$\Delta$DSP} & \multicolumn{2}{c}{Exact (\%)}\\
\cmidrule(lr){3-4}\cmidrule(lr){5-6}\cmidrule(l){7-8}
& {} & {Mean} & {SD} & \multicolumn{2}{c}{(pts)} & {Mean} & {SD}\\
\midrule
\multicolumn{8}{@{}l}{\textit{Training objective}}\\
Full, labels-only (final) & 21528 & 73.55 & 0.46 & $+$ & 12.47 & 67.07 & 0.65\\
Full, distilled & 21528 & 72.50 & 0.58 & $+$ & 11.42 & 64.97 & 0.76\\
\addlinespace[0.25ex]
\multicolumn{8}{@{}l}{\textit{Training recipe, labels-only}}\\
No curriculum & 21528 & 73.16 & 0.62 & $+$ & 12.08 & 66.07 & 0.56\\
\addlinespace[0.25ex]
\multicolumn{8}{@{}l}{\textit{Architecture, labels-only}}\\
Linear input & 21984 & 72.50 & 0.47 & $+$ & 11.42 & 65.13 & 0.87\\
Random-init $W$ & 21528 & 72.73 & 0.33 & $+$ & 11.65 & 65.33 & 0.69\\
Learn only $W$ & 576 & 60.25 & 0.45 & $-$ & 0.83 & 50.90 & 0.62\\
\bottomrule
\end{tabular*}
\end{table}

\subsection{SWD cross-corpus check}

We also score the selected labels-only artifact on the 24-track SC06
performance from the Schubert Winterreise Dataset~\cite{weiss2021swd}, chosen
because the release documents that performance as CC BY 3.0 audio. This is a
domain shift from FMAK/FMAKv2 popular and electronic material to a classical
song-cycle performance with piano and voice, so it tests whether the reranker
only fits the training-domain error profile. KL-FMAK reaches 93.33 MIREX and
91.67 exact, compared with 83.33/79.17 for KL-DSP and 87.08/83.33 for local
madmom. The paired MIREX delta over KL-DSP is $+10.00$ points with 95\% CI
$[-3.75,+25.00]$; the interval remains wide because the corpus is small.

The error mix narrows under reranking. KL-DSP has 19 exact matches, one fifth
error, one relative error, one parallel error, and two other errors. KL-FMAK
has 22 exact matches and two parallel errors, with no fifth, relative, or other
errors. Local madmom has 20 exact matches, one relative error, and three
parallel errors. This small check shows that the same artifact can correct some
non-harmonic DSP errors outside FMAK/FMAKv2, while the two remaining mistakes
are parallel-key confusions. A larger classical corpus is needed to confirm
corpus-level generalization.

\subsection{Runtime and deployment}

\begin{table}[t]
\centering
\caption{Deployment footprint for key estimation systems. Footprint
columns measure MB on disk after installation. Model is the learned model
artifact; binary/deps is the executable, library, runtime, and non-model
dependency closure. A zero means no learned model artifact.
Installed footprints compare deployable software ecosystems rather than
model-only parameter counts.
Rows keep KL-DSP and KL-FMAK adjacent, then list compact DSP tools and larger
learned systems.}
\label{tab:deploy}
\vspace{4pt}
\small
\setlength{\tabcolsep}{1.2pt}
\begin{tabular*}{\columnwidth}{@{\extracolsep{\fill}}l
  S[table-format=1.3]
  S[table-format=3.2]
  S[table-format=3.1]@{}}
\toprule
System & {\shortstack{Model\\(MB)}} & {\shortstack{Binary/deps\\(MB)}} &
  {\shortstack{Total\\(MB)}} \\
\midrule
KL-DSP & 0 & 2.0 & 2.0 \\
KL-FMAK & 0.115 & 76.58 & 76.7 \\
stratum-dsp & 0 & 1.4 & 1.4 \\
Essentia EDMM & 0 & 48.4 & 48.4 \\
libKeyFinder & 0 & 0.8 & 0.8 \\
madmom KeyCNN & 2.5 & 171 & 174 \\
S-KEY & 0.75 & 435 & 436 \\
\bottomrule
\end{tabular*}
\end{table}

\begin{figure}[!ht]
\centering
\includegraphics[width=.62\textwidth]{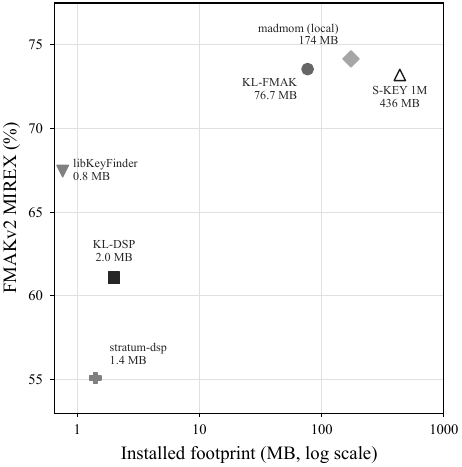}
\caption{FMAKv2 MIREX against installed footprint (Table~\ref{tab:deploy}).
Filled points use local measurements on the sealed holdout; libKeyFinder covers
996 scored tracks. The hollow S-KEY point combines a locally measured installed
footprint with a published full-protocol score. Essentia is omitted from the
scatter because it falls below the plotted y-range.}
\label{fig:footprint}
\end{figure}

\vspace{0.6ex}
For the footprint-accuracy comparison in Figure~\ref{fig:footprint}, local
FMAKv2 holdout scoring gives libKeyFinder 67.46 MIREX over 996 scored tracks,
stratum-dsp 55.09 MIREX, and Essentia EDMM 48.92 MIREX over all 1{,}000
tracks.
The KL-FMAK model artifact is a 115~KB fp32 ONNX file with 21{,}528
parameters. The 76.7~MB installed footprint in Table~\ref{tab:deploy} includes
the 2.0~MB trail mix binary, the 74.58~MB extracted macOS arm64 ONNX Runtime
installation, and the model artifact. The 0.32~ms figure measures
batch-size-one ONNX Runtime 1.30.0 CPUExecutionProvider inference over a
195-track FMAKv2 export fixture after one warmup track on a macOS arm64 Apple
M-series desktop, on top of the 486~ms DSP analysis on identical audio. This
fixture was drawn from the FMAKv2 holdout pool, but it is used only for the
runtime artifact, not for model selection or accuracy reporting. The selected
labels-only artifact uses the same exported graph and tensor shapes as that
latency artifact.
The ONNX Runtime footprint is therefore the measured evaluation package, not the
pure-Rust integration footprint. Removing the 74.58~MB ONNX Runtime term leaves
the 2.0~MB trail mix binary plus 0.115~MB of selected weights, about 2.1~MB
before measuring the final Rust application build.

\section{Discussion and Conclusion}

With a 115~KB ONNX model, KL-FMAK brings the DSP scorer within a point of
local madmom while adding 0.32~ms per track. It uses a much smaller model
artifact and a smaller installed footprint than the local madmom and S-KEY
baselines measured here, while keeping deterministic DSP as the base analyzer.
The frame-sequence branch is the supported source of the measured gain; the
DSP-margin curriculum, circular input, identity initialization, and distillation
show similar three-seed means and do not produce resolved gains in this sample.
We keep the circular input and identity-initialized DSP-score projection because
they encode transposition and DSP-score residual structure at negligible
parameter cost. Confirming the result on a broader cross-corpus evaluation,
beyond the single SWD check, is the main open item. Future work should also add
same-manifest measurements for additional learned baselines.

\unnumberedsection{Code availability}

The public implementation and paper artifact are available in the trail mix
repository:
\url{https://github.com/djpardis/trailmix/releases/tag/key-lime-paper-v1}.
The archived software release is available at
\url{https://doi.org/10.5281/zenodo.23866516}.
This tagged release records the code revision, selected model state, split
fingerprint, seeds, and evidence identifiers used for the reported results.

\unnumberedsection{Compliance with Ethical Standards}

This is a software and evaluation study for which no ethical
approval was required. Evaluation used publicly released datasets under
their original licenses: FMAKv2 annotations with audio from the Free Music
Archive, and the SC06 performance from the Schubert Winterreise Dataset. No new
data was collected.

\unnumberedsection{Conflict of Interest and Funding}

The author is affiliated with Cue Labs, which develops Cueport, a
local music application in the deployment setting discussed in the
introduction. No external funding was received for this work.

\bibliographystyle{IEEEbib}
\bibliography{references}

\end{document}